\documentclass[1p, review]{elsarticle}

\usepackage{lineno,hyperref}
\usepackage{booktabs}
\usepackage{wasysym}
\usepackage{amssymb}
\usepackage{ulem}
\usepackage{setspace}

\journal{Nuclear Instruments and Methods in Physics Research Section A}

\usepackage{siunitx}
\usepackage{listings}
\usepackage{xspace}
\usepackage{xcolor}

\newcommand{\htp}{\ensuremath{\mathrm{H}_2^+}\xspace}

\newcommand{\BE}[0]{\begin{equation}}
\newcommand{\EE}[0]{\end{equation}}
\newcommand{\BEA}[0]{\begin{eqnarray}}
\newcommand{\EEA}[0]{\end{eqnarray}}

\newcommand{\nuebar}{\ensuremath{\bar{\nu}_e}\xspace}

\mathchardef\mhyphen="2D

\newcommand{\libcom}{\textit{libcom1}\xspace}

\usepackage{enumitem}
\usepackage{booktabs}
\usepackage[frozencache,cachedir=.,newfloat]{minted}
\definecolor{LightGray}{gray}{0.9}
\definecolor{codegreen}{rgb}{0,0.6,0}
\definecolor{codegray}{rgb}{0.5,0.5,0.5}
\definecolor{codepurple}{rgb}{0.58,0,0.82}
\definecolor{backcolour}{rgb}{0.95,0.95,0.92}
\hypersetup{colorlinks = true,linkcolor = blue,anchorcolor =red,citecolor = blue,filecolor = red,urlcolor = red, pdfauthor=author}
\lstdefinestyle{mystyle}{
    backgroundcolor=\color{backcolour},   
    commentstyle=\color{codegreen},
    keywordstyle=\color{magenta},
    numberstyle=\tiny\color{codegray},
    stringstyle=\color{codepurple},
    basicstyle=\ttfamily\footnotesize,
    breakatwhitespace=false,         
    breaklines=true,                 
    captionpos=b,                    
    keepspaces=true,                 
    numbers=left,                    
    numbersep=5pt,                  
    showspaces=false,                
    showstringspaces=false,
    showtabs=false,                  
    tabsize=2
}

\begin{document}
\begin{frontmatter}

\title{The EPICS control system for IsoDAR}

\author{Philip Weigel\corref{mycorrespondingauthor}}
\cortext[mycorrespondingauthor]{Corresponding author}
\ead{pweigel@mit.edu}

\author{Monica Busza}
\author{Abutalib Namazov}
\author{Janette Park}
\author{Joshua Villarreal}
\author{Loyd H. Waites}
\author{Daniel Winklehner}

\address{Massachusetts Institute of Technology, 77 Massachusetts Ave, 
                     Cambridge, MA 02139, USA}

\begin{abstract}
Many large accelerator facilities have adopted the open-source EPICS software as the quasi-industry standard for control systems. They typically have access to their own electronics laboratory
and dedicated personnel for control system development. On the other hand,
small laboratories, many based at universities, use commercial software like LabView,
or entirely homebrewed systems. These often become cumbersome when the number of controlled devices increases over time.
Here we present a control system setup, 
based on a combination of EPICS, React Automation Studio, and
our own drivers for electronics available to smaller laboratories -- such as Arduinos --
that is flexible, modular, and robust. It allows small laboratories,
working with off-the-shelf modular electronics, power supplies,
and other devices to quickly set up a control system without a
large facility overhead, while retaining maximum compatibility and upgradeability. 
We demonstrate our setup for the MIST-1 ion source experiment at MIT.
This control system will later be used to serve the entire IsoDAR 
accelerator complex and, as such, must be easily expandable.
\end{abstract}



\end{frontmatter}


\section{Introduction}
\label{section:intro}
In the planned Isotope Decay-At-Rest experiment (IsoDAR)~\cite{bungau:isodar,Alonso:2022mup,alonso_neutrino_2022} 
we propose to search for sterile neutrinos through the disappearance of 
\nuebar, produced in a neutrino production target, and measured
in a nearby 2.6~kt scintillator detector at the Yemilab 
underground facility. In order to be decisive within a time-frame of 
five years, IsoDAR requires a continuous 60~MeV proton beam of 10~mA on target
at 80\,\% duty factor. We recently presented a mature design for 
a compact cyclotron capable of accelerating these record beam currents~\cite{winklehner_order--magnitude_2022} that we 
optimized using AI/ML techniques~\cite{koser_input_2022, villarreal_investigation_2022}. The design utilizes several novel ideas, 
among them RFQ direct injection~\cite{winklehner:rfq,WINKLEHNER2018231} and 
acceleration of \htp ions. 

We built a prototype for the ion source,
currently capable of delivering a 1~mA, \htp beam, at MIT~\cite{winklehner_high-current_2021-1,winklehner_new_2022}, which we are in the process of upgrading.
At the 10~mA nominal currents and in an underground laboratory setting, 
reliability and ease-of-use are paramount for a control system. 
Because we build and test
our hardware step-by-step, beginning with the ion source, then the RFQ, 
and finally the cyclotron, we also require an easily expandable control system.
Small laboratories, many based at universities, often use commercial software
like LabView~\cite{labview}, or entirely homebrewed systems. 
These often become cumbersome when the number of controlled devices 
increases over time. 
The Experimental Physics and Industrial Control
System (EPICS)~\cite{noauthor_epics_nodate} is the tool of choice for 
a multitude of large facilities around the globe, with applications in accelerators~\cite{poirier2016studies, uchiyama2017integration, Akeroyd_2018, SNS2022controls, Brodrick:2018mog}, fusion and neutron science~\cite{KIM20061829, cappelli2021status}, particle physics experiments~\cite{microboone2017design, mu2e2022controls, ritzert2015status}, astronomy~\cite{lupton2000keck, zhang2015telecontrols}, and more.

In this paper, we present the details of the implementation of a 
highly reliable and easily expandable framework based on 
EPICS, React Automation Studio (RAS)~\cite{ras,duckitt:ras2020}, and our own code
for 1) fast communication with microcontrollers and 2) data logging.
Our scripts and tutorials make this framework easy to deploy and 
to expand when needed. With just a personal computer or laptop 
and an Arduino, hardware control examples using EPICS can easily be set up 
and arbitrarily expanded later. This makes our setup ideal for educational purposes.
All our code is available freely on GitHub~\cite{winklehner_rfq-dip-epics_2022}.

The structure of this paper is as follows. 
In Section~\ref{section:controlsystem} we describe the various software packages,
in Section~\ref{section:implementation}, we discuss the implementation 
at our ion source laboratory, which doubles as an EPICS test-bench.
Finally, in Section~\ref{section:education} we discuss how our easy-to-use
and expandable setup can be deployed quickly to control small-scale 
experiments in labs, at home, or in an educational setting, by
presenting a set of instructions and tutorials.
Finally, in Section~\ref{section:conclusion} we present our conclusions.

\section{Control System Design}
\label{section:controlsystem}
\subsection{Devices and communications}
\label{section:libcom1}
In many small experiments, equipment is commonly controlled by off-the-shelf microcontrollers such as Arduinos or Raspberry Pi computers. 
While these excel at fast deployment and low maintenance, they are often not implemented efficiently in terms of communication with other devices. 
To remedy this, we have developed a common protocol for Arduino-based devices to standardize the serial communications of these devices. 
These devices can then talk directly to the EPICS Input/Output Controller (IOC) using a library compatible with this new protocol. 

\libcom is an improvement over a previous iteration of a communication library used for device communication with the IsoDAR Ion Source. 
This library offers faster processing times for the query/setting commands, while reducing storage and memory usage. 
The features are entirely implemented in C++ header files that can be imported into the Arduino library.
In the Arduino code, Channel objects are created with set/get functions and a ChannelMap object is defined with an array of Channel objects and their respective channel identifiers/numbers.
Querying a device with \libcom simply requires a serial command to be sent to the device in the format shown in  Table~\ref{table:libcom1}.
\begin{figure}
\centering
\begin{minted}
[
frame=lines,
framesep=2mm,
baselinestretch=1.2,
bgcolor=LightGray,
fontsize=\footnotesize,
% linenos
]
{C}
using mist1::com1::Channel;
MakeChannelMap(lookup, 3, ({
    Channel{ 't', 1, &get_temperature1, &dummySetFunc}, 
    Channel{ 'p', 1, &get_pressure1, &dummySetFunc},
    Channel{ 'p', 2, &get_pressure2, &dummySetFunc},
    Channel{ 'v', 1, &dummyGetFunc, &set_valve},
  })
);
\end{minted}
\caption{Example of creating a channel map object using \libcom on an Arduino microcontroller.}
\end{figure}

\renewcommand{\arraystretch}{1.5}
\begin{table}
    \centering
    \begin{tabular}{|c|c|c|}
        \hline
        Message Type & Format & Response \\
        \hline
        Query & q01[b][c][p] & o[b][c]$\pm$[m][e]$\pm$ \\
        \hline
        Query all & A & o\{[g]:\}$_{n}$ \\
        \hline
        Set & s[b][c][v] & None \\
        \hline
    \end{tabular}
    \caption{\libcom message types, formatting, and response. Note that ``Query All'' will have a response length that depends on the total number of channels $n$.}
    \label{table:libcom1}
\end{table}
The parameters in the formatting of messages in Table~\ref{table:libcom1} are given as:
\begin{itemize}[noitemsep]
    \item [b]: channel identifier
    \item [c]: channel number
    \item [p]: precision
    \item [m]: mantissa to \textless p\textgreater 
    \item [e]: absolute value of exponent
    \item [g]: \%g fprintf formatted float
    \item [v]: float value to set
\end{itemize}

Note that in this version of the library there are two modes for querying single devices: single channel and all channels.
The response given by an all channel query will concatenate $n$ values with a semi-colon for $n$ channels on the device.
Multiple queries per message may be supported by a future version of the library.

\subsection{EPICS}
\label{section:EPICS}
EPICS (Experimental Physics and Industrial Control System) is open source code developed at Argonne National Laboratory. 
EPICS is the quasi-industry standard for large particle accelerator systems. 
The main benefit of using EPICS is its scalability to very large systems with many devices, as it can support a wide range of protocols. 
This is ideal for the breadth of devices present in the \htp ion source. 
EPICS also offers real-time data communication and a high degree of user control.
The minimal latency makes it ideal for running controlled experiments.
The user control allows a user to set operating ranges for devices, create custom outputs, and create specific automated tasks, which makes it easy to interface with a Graphical User Interface (GUI).

In our control system design, EPICS serves as the bridge between the hardware and the user GUI. 
Many of the devices are in-house electronics that utilize off-the-shelf parts, and use \libcom to communicate with EPICS.
EPICS process variables (PVs) are used to define the types of inputs, outputs, and calculations of the control system and their properties.
The GUI, discussed further in Section~\ref{section:RAS}, interacts with the EPICS PVs through EPICS Channel Access (CA).

\subsubsection{libcom1 Integration}
\label{section:EPICS:libcom1}
Integrating \libcom into the EPICS framework requires a minor modification to the \textit{StreamDevice} package. 
This modification is simply an extension to the format converters to correctly parse the incoming data from devices using \libcom.
Once compiled with EPICS, the formatting tools can be used in the protocol files.
\begin{figure}
\centering
\begin{minted}
[
frame=lines,
framesep=2mm,
baselinestretch=1.2,
bgcolor=LightGray,
fontsize=\footnotesize,
]
{C}
# Get a value from a channel
com1_getLink {
    # q01<chanelnum><precision>
    out "q01\$1\$2";
    # o<channelnum><com1exponent>
    in "o%*c%*c%z";
}

# Set a value on a channel
com1_setLink {
    # s<channelnum><value>
    out "s\$1%d";
}
# Read all channels
read_all_array {
    separator = ":";
    extrainput = ignore;
    out "A";
    in "o%g"
}
\end{minted}
\caption{Example EPICS protocol file for \libcom}
\end{figure}
\subsection{React Automation Studio}
\label{section:RAS}
React Automation Studio (RAS) \cite{ras, duckitt:ras2020} is a scalable platform for instrumentation control which utilizes containerized services to provide EPICS integration, logging, alarm handling, a React GUI front end, and more. 
The suite can be deployed on a machine and access the lab devices through the local network while exposing the control system interface to the internet with a secure login, allowing for remote control and monitoring if required. 
Alarm handlers are linked to EPICS alarms which can alert operators through front-end notifications and email, and are logged through a MongoDB service. 
An important feature is the ability to save and load set points easily to restore components into a predetermined state for operation. 
These values are obtained from the EPICS PVs and stored in a separate MongoDB database and its functionalities can be accessed through standard system components within RAS. 

RAS uses Docker, meaning the aforementioned containers serve as virtual machines that are initialized with the required environments on any suitable machine (currently, Linux is supported). 
The following two subsections will briefly explain the end-to-end integration and high-level architecture of RAS.

\subsubsection{RAS Front end --- React}
\label{section:RAS:React}
Since RAS can be deployed on the web, it can be used on any device with an online or local connection to the server. 
The \textit{React} framework is used as it allows flexible development while offering a reactive and real-time experience to the user. 
A responsive interface based on \textit{Material-UI} is implemented to serve a clean GUI on both desktop and mobile devices. 
Reusable components in the web views provide a high degree of flexibility for designing a control system, such as displays for a wide range of instrumentation.
These components take one or more PVs as input (which can be declared using macros for convenience) and instantiate themselves using relevant queried data. 
The front-end queries the pvServer using the \textit{Socket.IO} socket library in real time and posts user input data to the APIs offered by pvServer. 
Plots are generated using the \textit{Plotly} graphic library, which offers customizable views with variety of data visualization tools. A few additional tools for visualization were developed for this work, which are available in our GitHub repository
\cite{winklehner_rfq-dip-epics_2022}.

The RAS front end also comes with useful configuration pages, including the management of macros, access rights, and alarm handling. 
The main administrator user can set access rights to specific PVs or other configurations (such as alarm handler settings) for other users. 
Alarms can be set in the UI, targeting values from given PV to watch and send notifications when those values violate the alarm conditions.
These notifications are displayed on the interface and can be sent through messaging services or email.
All alarms are logged and the users are able to view the history of alarms and query for specific targets.

The development of the front end of RAS requires basic \textit{React}, \textit{HTML}, \textit{CSS}, and \textit{JavaScript} knowledge. 
However, due to the modularity of the implementation, most developers can use the existing components as black boxes and build on top of them conveniently.

\subsubsection{RAS Back end --- Python}
\label{section:RAS:python}
The pvServer is the \textit{Python} back-end service for RAS using \textit{Flask} framework, which establishes connections to the EPICS PVs. 
EPICS is integrated into the RAS pvServer using \textit{PyEPICS} Channel Access (CA). 
\textit{Flask-SocketIO} is used to create sockets for sending real-time updates to the front end. 
This connection only happens when the user is authenticated and have access rights to the requested PV, and otherwise they are denied the connection. 
This back end also handles the user authentication, permissions, and provides APIs for writes to EPICS variables, which are handled similarly.
\textit{MongoDB} databases are used to manage some forms of data storage and alarm handling.
Alarm configurations can be setup in \textit{JSON} files, allowing for flexible conditions and predetermined alarm areas.
Previous alarms that were triggered can be accessed at a later date through the alarm logs.

\subsection{Data Logger}
\label{section:datalogger}
A new data logging feature was developed to store real-time data fast and efficiently. 
An important design principle was to avoid any changes to the existing RAS framework to simply integration. 
The data logger interfaces with the pvServer by creating new \textit{SocketIO} clients for each monitored PV. 
The clients wait for any data to be transmitted and store it in a buffer. Once the buffer is full, the data contained within it will be saved as a chunk to a file.
The default file format is HDF5, but additional file types can be implemented within the socket methods of the logger file.

The data logger is initialized with a configuration file, which allows the user to specify networking and data storage information as well as a list of desired PVs to monitor and their data types. 
To accommodate the varying rates at which certain PVs need to be logged, two scanning modes have been added: sampling and continuous. 
In the sampling mode, the user can specify a rate to poll the PV. In continuous mode, data will only be logged when the PV is updated by an EPICS IOC. 
Specifying the sample rates for individual devices reduces storage space usage by reducing unnecessary polling. 
The continuous mode is useful for storing values that don't change often, such as an interlock or a switch.
\begin{figure}
\centering
\begin{minted}
[
frame=lines,
framesep=2mm,
baselinestretch=1.2,
bgcolor=LightGray,
fontsize=\footnotesize,
% linenos
]
{C}
# Data logger example configuration
{  
  # pvServer Settings
  "server_address": 127.0.0.1,
  "server_port": 9001,
  "server_namespace": "/pvServer",

  # Logging settings
  "buffer_size": 64,
  "data_path": "./data/",

  # PVs to log
  "pvs" : {
    "device1:switch" : {
      "value_dtype": "bool",
      "time_dtype": "f8",
      "scan_type": "continuous",
      "sample_rate":
    },
    "device2:pressure" : {
      "value_dtype": "f8",
      "time_dtype": "f8",
      "scan_type": "sample",
      "sample_rate": 1,
    },
  }
}
\end{minted}
\caption{An example configuration file for ras\_datalogger.}
\end{figure}
The data logger is contained within a simple python script and requires few dependencies. 
This lightweight design allows for quick deployment and the ability to have multiple devices on the network storing data without significant overhead.
The code for this tool is available on GitHub~\cite{weigel:ras_datalogger}.

\section{Example: Simple Control System}
\label{section:simple_example}
\begin{figure}
    \centering
    \centerline{\includegraphics[width=0.5\columnwidth]{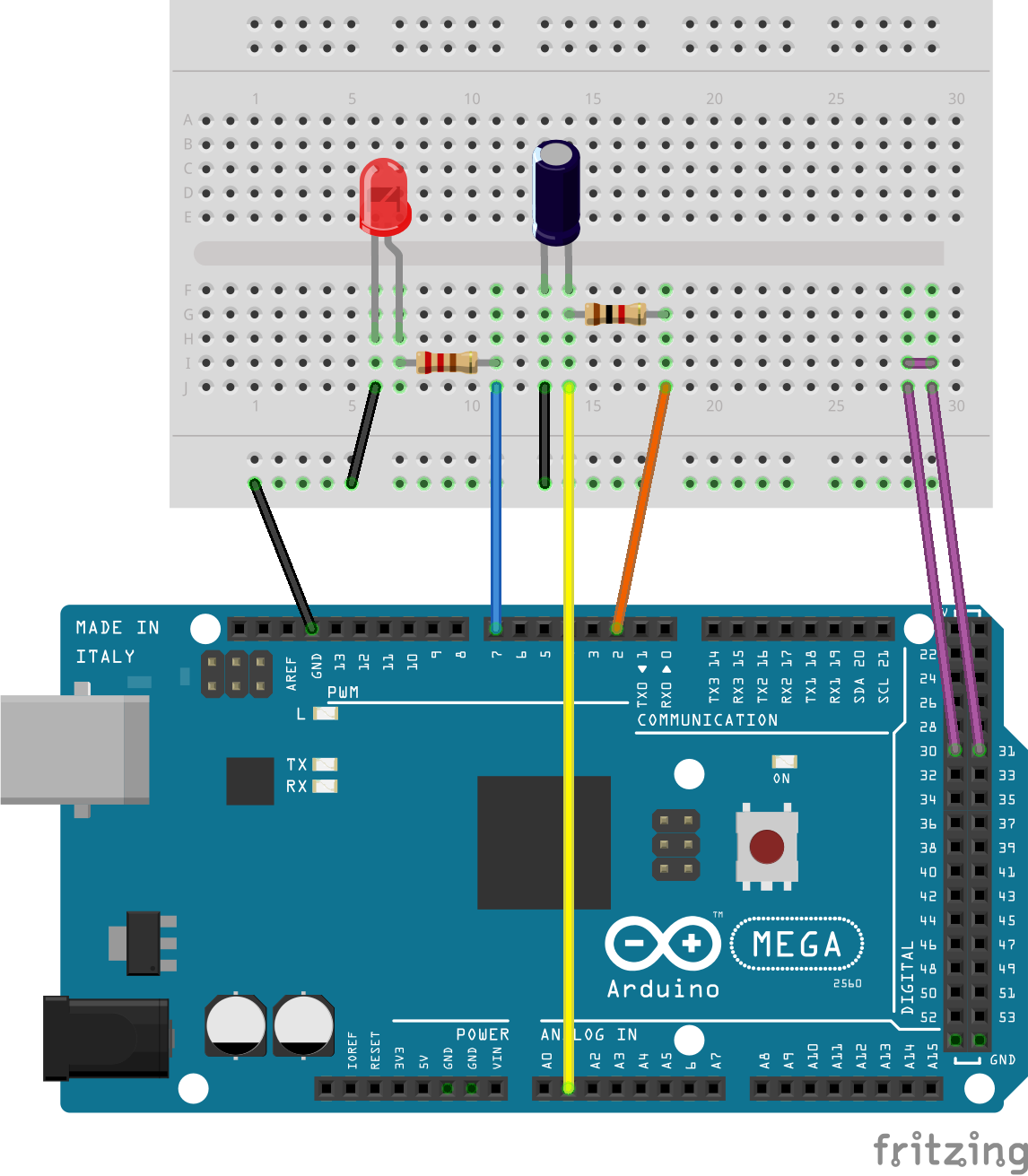}}
    \caption{Example Arduino setup for testing and training. Here, DIO~7 is configured as output 
             to switch on/off an LED. A 220~$\Omega$ resistor protects the LED from over-current.
             DIO~2 is configured as PWM output to generate 0-5 V via pulse-width modulation.
             The output is smoothed with a low-pass RC circuit (1~k$\Omega$, 0.1~mF) and 
             piped back to AI~1. Finally, DIO~30 is configured as output and DIO~31 
             is configured as input and they are interconnected.
             The diagram was produced with Fritzing.org~\cite{fritzing_web}.}
    \label{fig:example_arduino}
\end{figure}
Included in the software repository is a simple working example of all the components described above. This simple example demonstrates how to map inputs and outputs of an Arduino microcontroller to EPICS with \libcom, and a bare-bones interface using RAS. The purpose of this is to give first-time users a straightforward, easy to test example to get familiar with the individual components before building a more complex application. The default Arduino configuration assumes a few simple connections between input and output pins, as seen in Figure~\ref{fig:example_arduino}. The Arduino sketch file contains the code for controlling these pins and creating the \libcom channels. The EPICS database files simply create input and output records for these pins, requiring no modification assuming the default layout. Finally, there is a folder that includes a few files for RAS that create a page with buttons for toggling the outputs, indicators for the binary inputs, and a plot for the analog inputs. More complex behavior can be achieved with simple modifications to the individual components described here.\\

\begin{figure}
    \centering
    \centerline{\includegraphics[width=0.425\columnwidth]{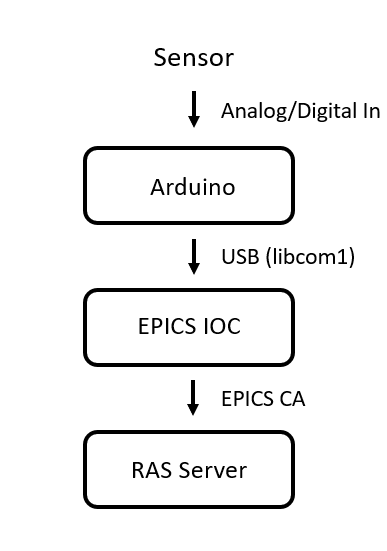}}
    \caption{An example of a set up for a generic sensor. An analog or digital signal is read by an Arduino running a program that maps the input to a \libcom channel. An EPICS IOC with the \libcom protocol can read values from the Arduino over USB. Lastly, the pvServer within RAS communicates with the IOC via EPICS CA to display the value.}
    \label{fig:example_flowchart}
\end{figure}

\section{Example: Ion Source Control System}
\label{section:implementation}

\renewcommand{\arraystretch}{1.5}
\begin{table*} 
    \centering
    \resizebox{\textwidth}{!}{%
    \begin{tabular}{|c|c|c|}
        \hline
        Device(s) & Driver Type & Comments \\
        \hline
        Matsusada AU20P7 power supplies & Matsusada CO-series drivers & High voltage  \\
        \hline
        Ion gauge controller & libcom1 & In-house electronics  \\
        \hline
        Thermocouples & '' & In-house electronics \\
        \hline
        Interlocks & '' & In-house electronics \\
        \hline
        Water flow meters & '' & In-house electronics \\
        \hline
        Vacuum system valves & '' & In-house electronics \\
        \hline
        Matsusada REK 24-150 power Supply & Generic FTDI USB-to-RS232 & Filament discharge  \\
        \hline
        TDK GEN8-300 power supply & TDK D2XX driver (FTDI) & Filament heating \\
        \hline
        MKS GV50A mass flow controller & Generic FTDI USB-to-RS485 & H$_2$ gas inlet \\
        \hline
    \end{tabular}}
    \caption{Devices currently employed in the MIT Ion Source Lab and controlled
            by EPICS/RAS.}
    \label{table:devices}
\end{table*}
\begin{figure*}
    \centering
    \includegraphics[width=\textwidth]{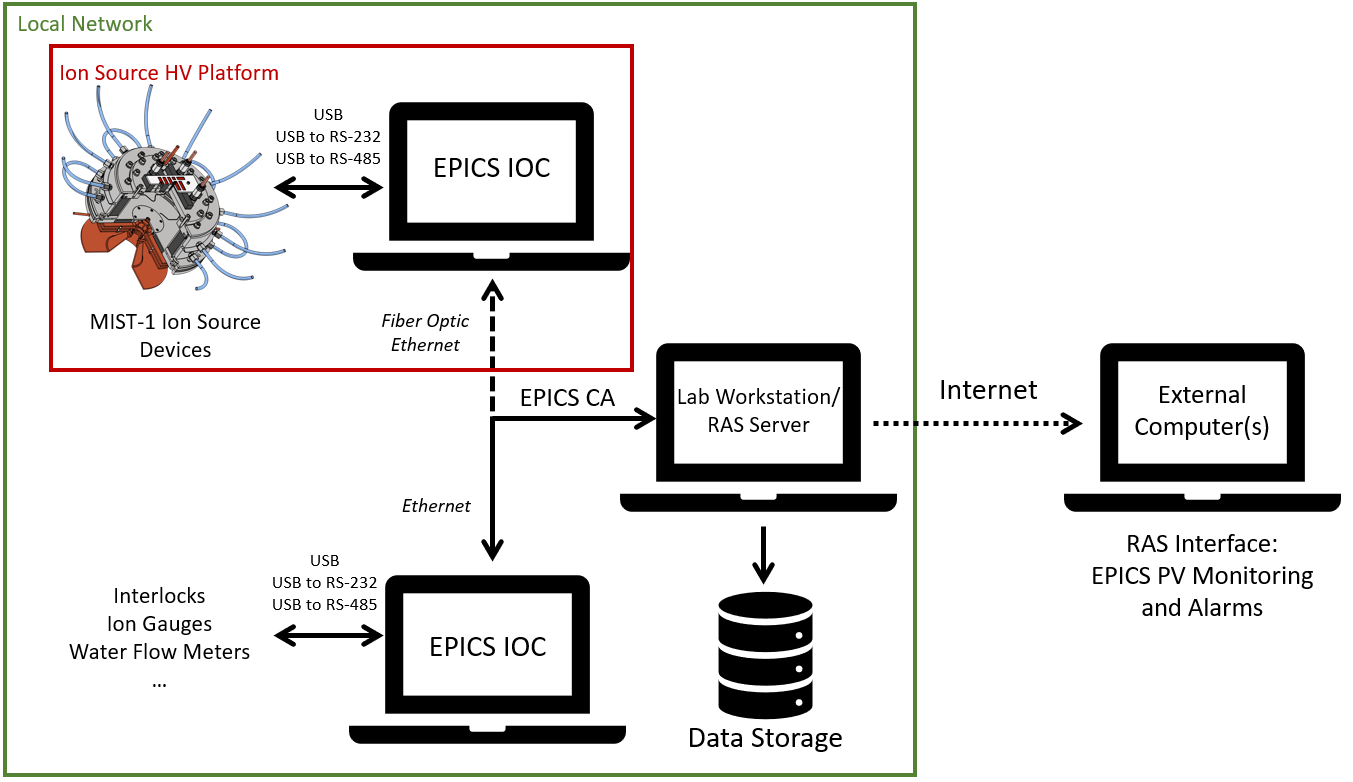}
    \caption{The layout of the MIST-1 \htp Ion Source control system. The RAS server is set up to run on the lab workstation and serves as the outward/internet-facing device. Two EPICS IOCs are used for the controls of different devices. One is on the high voltage platform and is used to interface with the power supplies that drive the source. A fiber optic Ethernet adapter is used to provide electrical isolation from the high voltage. The other IOC handles other devices off of the source platform, such as the interlocks. The data-logging service is also run on the lab workstation, but is implemented such that it can be deployed on any computer on the local network.}
    \label{fig:ion_source}
\end{figure*}
The IsoDAR MIST-1 \htp ion source uses the tools highlighted in the previous sections to create a stable and easy-to-use control system. 
Many of the devices utilize custom electronics and use microcontroller (e.g. Arduino
and Teensy) to read in or output voltages.
Communication with these devices is streamlined using the \libcom protocol over USB, which allows for a standardized and fast way of integrating new devices into the control system.
In this example, two EPICS IOCs are used: one to communicate with devices on the high voltage platform inside the field cage of the ion source, and another for all devices outside.
An optical fiber Ethernet cable is used for communicating with the platform IOC for electrical isolation.
The packets are first decoded using a protocol file, which employs \textit{StreamDevice} discussed in Section~\ref{section:EPICS:libcom1}.
Other devices which require manufacturer-supplied drivers or serial protocols can also be interfaced with EPICS as stream devices.
Each device type has an associated database files that contains all relevant records and application-specific alarms.
For example, a power supply will contain records to get/set/read voltages and currents with thresholds set for over-voltage and over-current alarms.
An example of the process variables defined for the Matusada AU20P7 power supply is shown in Table~\ref{table:matsusada_pvs}.
Another computer on the local network runs the docker containers for RAS suite, which also serves as the control computer for operators.
The RAS server runs the docker containers within a Windows Subsystem for Linux (WSL) instance running Ubuntu 20.04 and the front end of the control system can be accessed from any web browser on the local network.
\begin{table*}
    \centering
    \begin{tabular}{|c|c|c|c|}
        \hline
        Process Variable & Description & Process Variable & Description \\
        \hline
        voltage:set & Set output voltage & polarity:set & Set output polarity \\
        \hline
        voltage:get & Get last set voltage & polarity:get & Get last set polarity \\
        \hline
        voltage:read & Read output voltage & polarity:read & Read output polarity \\
        \hline
        current:set & Set output current & status:set & Set output status \\
        \hline
        current:get & Get last set current & status:read & Read output status \\
        \hline
        current:read & Read output current & reset & Reset cut-off state \\
        \hline
    \end{tabular}
    \caption{Process variables associated with the Matusada AU-20P7 power supply. Additional commands such as enabling/disable remote control are also available through the Matsusada communication protocol, but are not currently implemented.}
    \label{table:matsusada_pvs}
\end{table*}

Operation of the control system requires careful monitoring of the voltages and currents of various power supplies over time, so the MIST-1 ion source pages use a host of different plotting features from existing RAS libraries as well as new application-specific modifications to existing tools.
React components have also been created for certain device types, which allows for modular development of the beam line controls.

There are two methods for saving values from the PVs in the control system. 
The first is by using the native RAS tools, which are convenient for saving set points and other values of interest. 
This method saves the data to a MongoDB database running on the RAS server, which can be accessed by other computers on the local network.
The other is using the data logger described in Section~\ref{section:datalogger}, which can be used to save data from devices that can change rapidly (i.e. a Faraday cup or emittance scanner).
The data stored in the HDF5 files can easily be accessed from analysis scripts later.

\subsection{Safety}
\label{section:safety}
The ion source requires several mechanisms to ensure safe operations, the most important of which is preventing access to the high voltage areas.
Interlocks have been implemented for the two access doors to the field cage, either of which will power down all power supplies upon opening.
The door switches to the field cage are connected to a microcontroller which is scanned by EPICS for changes.
One such application of this interlock scheme with the Matsusada REK power supply is shown in~\ref{table:interlock_example}.
When one of the switches is released, EPICS will trigger an alarm and disable the Matsusada high voltage power supplies as well as the filament heating and discharge power supplies.
The EPICS alarm will trigger an alarm in RAS to notify operators and log the event.
RAS also has the ability to send alerts via email or SMS when certain alarms are triggered.
Monitoring the temperature of the source body is also important for operations, as exceedingly high temperatures can permanently damage the Sm$_{2}$Co$_{17}$ magnets that produce the magnetic containment field within the chamber.
To prevent this, water flow meters that monitor the source cooling water and thermocouples on the source body itself are routinely scanned for changes.
Another heating concern is the melting of the filament, which can  occur if there is a loss of vacuum, and so there is an upper allowed limit on the source pressure during running.
If changes to these variables go outside of the normal operating conditions, the source heating power supply is disabled and an alarm is triggered.

\begin{table*}
    \centering
    \resizebox{\textwidth}{!}{%
    \begin{tabular}{|c|c|c|c|c|}
        \hline
        Process Variable & Input A & Input B & Output & Description \\
        \hline
        i1:get\_interlock & & & & Input record for the first interlock \\
        \hline
        i2:get\_interlock & & & & Input record for the second interlock \\
        \hline
        interlock\_status & i1:get\_interlock & i2:get\_interlock & A OR B & Logical OR of the two interlock statuses \\
        \hline
        REK:SetToggleOutput & & & & Input record for user-controlled output status \\
        \hline
        REK:OutputIntlk & interlock\_status & REK:SetToggleOutput & A ? 0 : B & Disables output status if the interlock triggers \\
        \hline
        REK:ToggleOutput & REK:OutputIntlk & & & Enables/disables output based on the value of REK:OutputIntlk \\
        \hline
    \end{tabular}
    }
    \caption{An example of an interlock setup in which there are two interlocks safeguarding access to a high voltage device: i1:get\_interlock and i2:get\_interlock. If either interlock is triggered, a global interlock status flag (interlock\_status) is set high. A process variable specific to the power supply, REK:OutputIntlk, will set its output to 0 if interlock\_status is high, or to the value of REK:SetToggleOutput (set by operators in the control system) if it is low. Lastly, the process variable REK:ToggleOutput is used to send a command to the power supply to set its output status.}
    \label{table:interlock_example}
\end{table*}

\section{Test Measurement}
\label{sec:firsttest}

\begin{figure}[t!]
    \centering
    \includegraphics[width=0.5\textwidth]{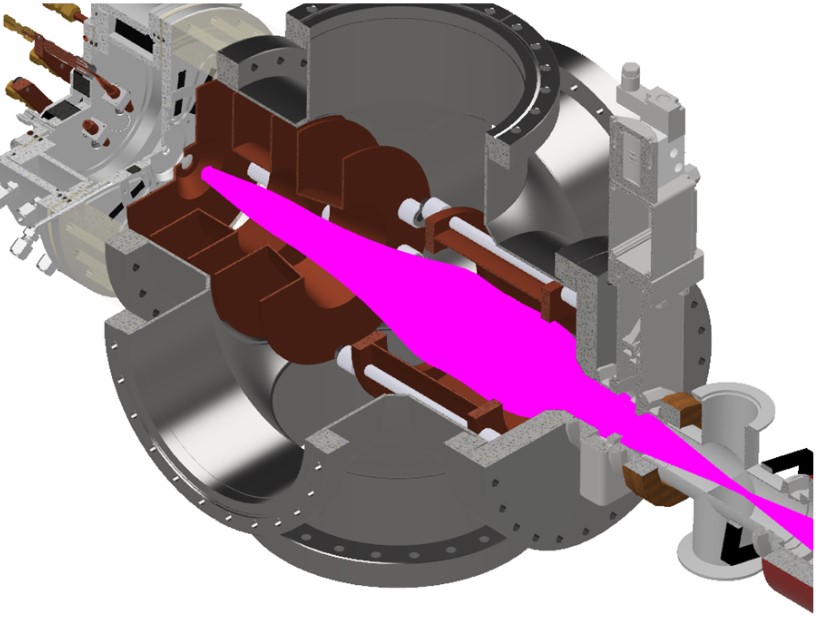}
    \caption{CAD rendering of the new extraction system and LEBT for the MIST-1 ion
             source that was installed in parallel with the upgrade to EPICS. Typical voltages
             are listed in Table~\ref{tab:test_settings}. Figure from Ref.~\cite{waitesLowEnergyBeam2022}.}
    \label{fig:xs}
\end{figure}

\begin{figure*}[t!]
    \centering
    \includegraphics[width=\textwidth]{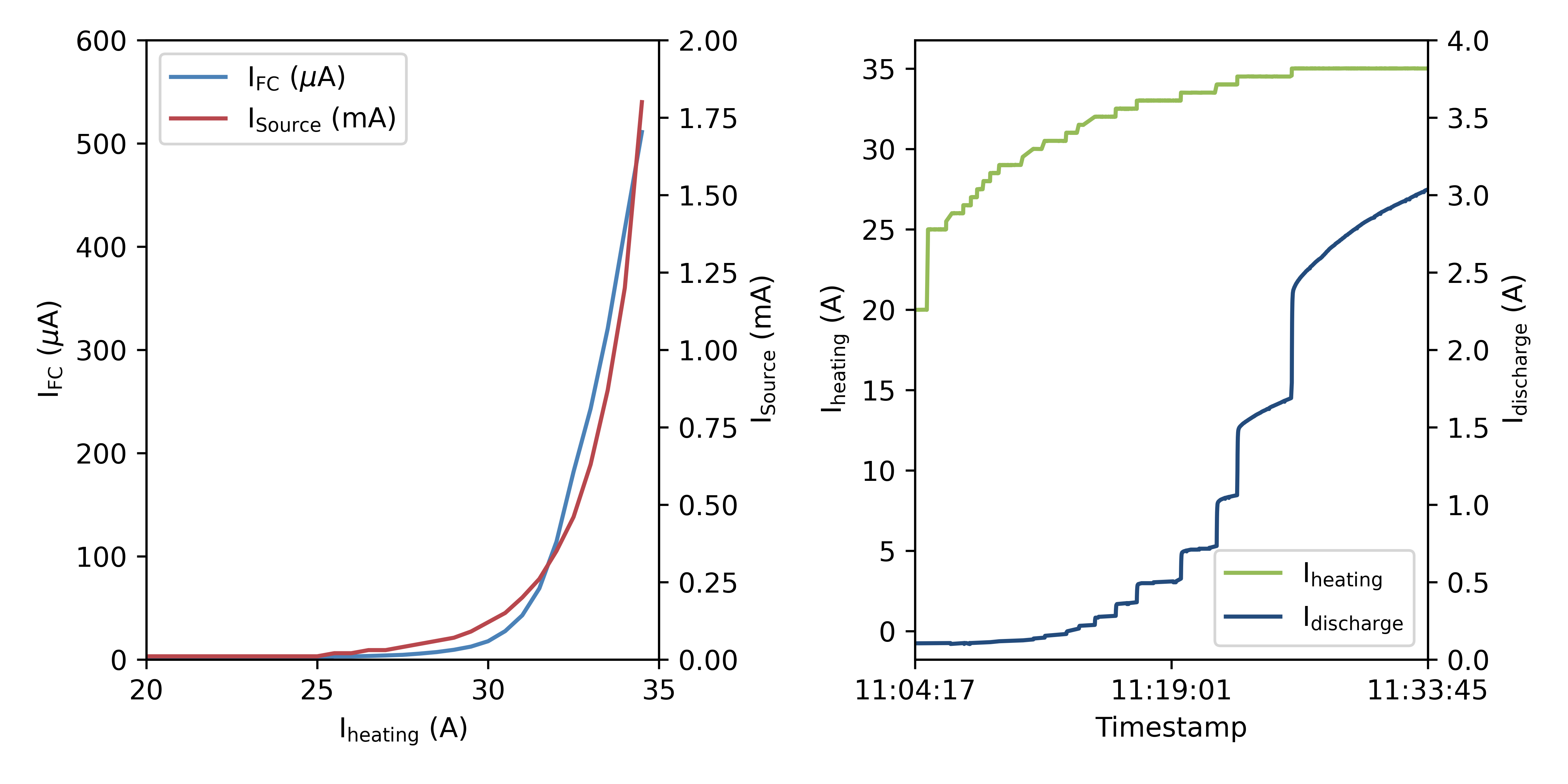}
    \caption{Left: Drain current on the high-voltage power supply of the ion source
             and current measured in the Faraday cup as functions of filament heating
             current. Right: Filament heating current and plasma discharge current time
             series as recorded by the data logger.}
    \label{fig:measurement}
\end{figure*}

\renewcommand{\arraystretch}{1.5}
\begin{table} [b!]
    \centering
    \begin{tabular}{|l|l|l|l|}
        \hline
        Device\hspace{20pt} & Setting \hspace{20pt} & Device \hspace{20pt} & Setting \hspace{20pt} \\
        \hline
        V$_\mathrm{source}$ & 15~kV & MFC & 15\% (0.75 sccm)\\
        V$_\mathrm{puller}$ & -2~kV & V$_\mathrm{discharge}$ & 100~V\\
        V$_\mathrm{lens1}$ & 10~kV  & P$_\mathrm{chamber1}$& $9.4\cdot10^{-6}$~Torr\\
        V$_\mathrm{lens2}$ & -5~kV  & &\\
        V$_\mathrm{lens3}$ & 0~kV  & &\\
        \hline
    \end{tabular}
    \caption{Settings during the filament heating test.}
    \label{tab:test_settings}
\end{table}

Here, we present a test measurement to demonstrate the functionality of the 
EPICS-based control system for the MIST-1 ion source. This measurement
was part of the first commissioning runs after switching to EPICS and after 
the installation of a new pentode electrostatic extraction 
system~\cite{waitesLowEnergyBeam2022} (cf. Figure~\ref{fig:xs}). 
During  this test, the filament heating current was 
varied from 20~A to 35~A. The other settings were kept constant and are listed in
Table~\ref{tab:test_settings}. It should be noted that normally, with increasing beam
current, the voltages on the extraction system electrodes would be adjusted 
to ensure optimal transport. However, in this case, because the resulting beam current
only goes up to 0.5~mA during this early test and the Faraday cup sits right after
the last lens element, this was not needed. We show the extracted beam current 
together with the source drain current as a function of filament heating in
Figure~\ref{fig:measurement}, left.
We recorded the source parameters in hdf5 files using the data logger 
and subsequently plotted them in Figure~\ref{fig:measurement}, right.
The control system performed as expected.

\section{Educational Use}
\label{section:education}
The containerized control system as described here provides an excellent base for small systems in educational labs that require communication with off-the-shelf micro-controllers and 
RS232-, RS485-, or USB-enabled power supplies and other devices. 
The GitHub repository~\cite{winklehner_rfq-dip-epics_2022} for this work contains instructions for setting up and deploying a bare-bones version of EPICS and RAS, with additional code for the \libcom library and the corresponding \textit{StreamDevice} modification for EPICS. 
The individual tools described here may require more time to develop familiarity with, 
but the workflow of connecting a serial device and creating display elements is simple. 
The provided example (see Figure~\ref{fig:example_arduino})
only requires an Arduino, a few readily available components,
and a platform to run EPICS and RAS, e.g., a low-cost laptop.
The preferred operating system is Linux, but instructions using 
the Windows Subsystem for Linux (WSL-2) are available too.

\section{Conclusion}
\label{section:conclusion}
In this paper, we presented the control system for the MIST-1 
ion source, which is based on a combination of EPICS, React Automation Studio,
and code we wrote for serial communication with Arduinos.
We also introduced a tutorial and a set of instructions and examples
to ease the installation of similar control systems. These
are freely available on GitHub.
We find that our setup is reliable and, due to the modular nature
of EPICS and RAS, easily expandable, which ideally suits our needs
to slowly expand the system. As the next steps, we will 
include a radiofrequency quadrupole linear accelerator 
and later a full cyclotron particle accelerator. The GitHub 
repository and tutorials we generated are well-suited for educational
purposes, e.g., lab-courses, or DIY projects.

\section{Acknowledgements}
The authors are very thankful for the contributions of Aashish Tripathee 
to a previous version of the code for Arduino communication and 
Ryan Yang to \libcom and the first setup of the EPICS back-end for MIST-1.
This work was supported by NSF grants PHY-1505858 and 
PHY-1626069, as well as the Heising-Simons Foundation.
D. Winklehner was supported by funding from the Bose Foundation.
The authors would like to thank Janet Conrad for her support.

\bibliography{epics.bib}

\end{document}